\documentclass[reprint, amsmath, amssymb, aps, nofootinbib]{revtex4-2}

\usepackage{graphicx}
\usepackage{dcolumn}
\usepackage{bm}
\usepackage{gensymb}
\usepackage{color}
\usepackage{hyperref}
\usepackage{fancyvrb}

\definecolor{blue}{rgb}{0.1,0.3,1}
\definecolor{green}{rgb}{0.1,0.6,0.1}
\definecolor{red}{rgb}{1,0,0}
\definecolor{pink}{rgb}{0.9,0.3,0.7}
\definecolor{azur}{rgb}{0,0.5,0.5}
\definecolor{orange}{rgb}{1,0.5,0.2}
\definecolor{brown}{rgb}{0.5,0,0}

\newcommand{\ie}{\textit{i.e.~}}

\DeclareMathOperator*{\argmax}{arg\,max}

\begin{document}

% \preprint{APS/123-QED}

\title{Maze-solving with density-driven swarms}

\author{Esther María Zamora Sánchez}
\author{Sébastien Billès}
\author{Paul-Henry Glinel}
\author{Nicolas Bredeche}
 \altaffiliation[Also at ]{Sorbonne Universit\'{e}, CNRS, Institut des Syst\`{e}mes Intelligents et de Robotique, ISIR, F-75005 Paris, France}
\author{Raphaël Candelier}
 \email{raphael.candelier@sorbonne-universite.fr}
\affiliation{Sorbonne Université, CNRS, Institut de Biologie Paris-Seine (IBPS), Laboratoire Jean Perrin (LJP), F-75005, Paris}

\date{\today}

\begin{abstract}
We propose a new kind of collective motion where swarms of simple agents are able to find and fix the solution of two-dimensional mazes. The model consists of active memoryless particles interacting exclusively via short-ranged perception of local density and orientations. This system generates traveling density waves when constrained in one dimension, and self-organized swarms exploring local branches in two-dimensional mazes. Depending on a single kinetic parameter, the swarms can develop large tails and further gain long-term persistence, which ultimately allows them to robustly solve mazes of virtually any kind and size. By systematic exploration of the parameter space, we show that there exists a fast solving regime where the resolution time is linear in number of squares, hence making it an efficient maze-solving algorithm. Our model represents a new class of active systems with unprecedented contrast between the minimality of the processed information and the complexity of the resolved task, which is of prime importance for the interpretation and modeling of collective intelligence in living systems as well as for the design of future swarms of active particles and robots.
\end{abstract}

\maketitle

Swarm intelligence is the collective behavior that allows a self-organized population to succeed at solving a task where single individuals would fail. In the absence of centralized control dictating how individual agents should behave, it is the local interactions among agents and with the environment that lead to the emergence of a global behavior, unknown to the agents~\cite{Bonabeau_1999,Krause_2010}. Examples of swarm intelligence in natural systems are abundant, including ant~\cite{Bega_2019} and bee colonies~\cite{Lemanski_2019}, bird flocks~\cite{Ballerini_2008}, fish schools~\cite{Partridge_1982} and microbial intelligence~\cite{Westerhoff_2014}, and the concept has inspired many other fields like soft matter~\cite{Hamann_2018,Baulin_2025}, swarm robotics~\cite{Kennedy_1995,Duan_2023} and optimization~\cite{Priyadarshi_2025}.

\begin{figure}[b!]
\centering
\includegraphics{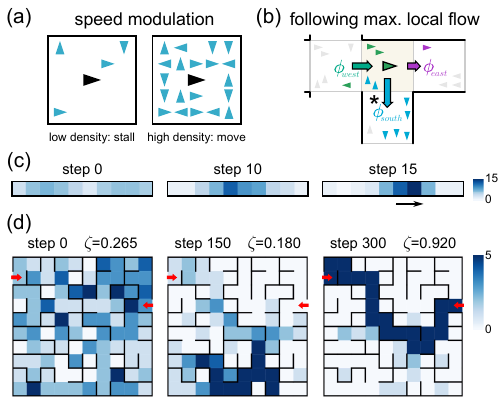}
\caption{Agents dynamics. 
(a-b) Scheme of the two kinetic rules. In (b), the local outward flows are $\phi_{west}=-2$, $\phi_{east}=+1$ and $\phi_{south}=+3$, so moving agents would go south. Agents in grey have irrelevant orientations and are ignored. 
(c) Dynamics of 50 agents with $\eta=10$ in a closed corridor of $s=10$ squares. Initial positions are random, and a soliton rapidly forms. Color indicates the number of agents on each square.
(d) Solving of a square maze of size $a=10$ by 200 agents with $\eta=25$, with random initial positions and orientations. $\zeta$ is the solving ratio. The full dynamics is available in movie 1.}
\label{fig_1}
\end{figure}

Despite this large corpus of systems and applications, it is still unclear what minimal set of features in the interactions is sufficient to produce swarm intelligence. For instance, navigation in unknown, constrained environments is a paramount challenge for both natural and robotic swarms; for such tasks, most of the documented behaviors rely on one or a combination of the three following strategies: (\textit{i}) memory keeping by the agents, often in the form of paths or maps, (\textit{ii}) explicit communication of the acquired information to the group and (\textit{iii}) stigmergy, \ie indirect coordination through the environment, like the pheromones used by more than $7,000$ insect species~\cite{Symonds_2008}.

In this letter, we revisit the classical navigation task of two-dimensional maze solving and demonstrate that none of the aforementioned strategies are actually necessary. In the spirit of the minimalist models of collective motion~\cite{Ouellette_2022} and rule-based navigation behavior, we used memoryless, communicationless and markerless agents that have only access to the local density and orientation of their neighbors to take decisions. We show that with just two simple but non-linear kinetic rules governing respectively the probability and the orientation of motion, a collection of agents constrained in a corridor self-organize to create density solitons. In two-dimensional mazes, such swarms display both short-term and long-term persistence at intersections that give rise to, respectively, exploration and exploitation capabilities. This ultimately allows them to collectively solve a task as difficult as finding and fixing the solution of mazes of any kind and size. We further show that there exists a fast solving regime in the parameter space where the resolution time is linear in maze complexity.\\

% \section*{\label{sec:level1}Simulations and results}
\label{sec:level1}\textbf{Simulations and results.} Let us consider a discrete, two-dimensional square mesh comprising $s$ squares where a total of $N$ agents move from square to square. We define $n_i^k$ the number of agents on square $i$ coming from square $k$, the local density as the total number of agents on a square $n_i = \sum_{k} n_i^k$ and the global density as the average number of agents per square $d=\sum_{i} n_i/s=N/s$. Each agent has only access to the number of agents $n_i^k$ for all orientations on their own square and on the direct adjacent accessible squares. To decide their next action, agents follow two kinetic rules (fig.~\ref{fig_1}-a,b):
\begin{enumerate}
  \item The moving probability for an agent on square $i$ is
  \begin{equation}
    p_{move} = \frac{n_i}{n_i + \eta}
    \label{eq:p_move}
  \end{equation}
  \noindent where $\eta$ is refered to as the \textit{kinetic parameter} in the following.
  \item When an agent moves, it follows the maximal outward flow at its current position $i$, defined as:
  \begin{equation}
    \argmax_{j \in J(i)}(\phi_j) \quad \text{where} \quad  \phi_j = n_j^i - n_i^j
    \label{eq:flow}
  \end{equation}
  \noindent and $J(i)$ is the set of accessible squares next to $i$. Agents count themselves in the relevant $n_i^j$, and in case of a tie each moving agent follows at random the direction of one of the maximal flows.
\end{enumerate}

To avoid a bias due to single-square dead ends in the computation of flows at intersections, we also implemented that agents arriving at a dead-end square immediately flip their orientation; this removes temporary geometrical trappings and speeds up the overall solving process.

Note that all agents on a given square perceive the exact same inputs, so the system is directly transposable to a cellular automaton where cells could store sets of tokens and pass them to their neighbors at each iteration with mass conservation of mass. As we are more interested in agent-based models, we will stick to this view in the sequel.\\

% \subsection*{Behavior in corridors}
\label{sec:corridors}\textit{Behavior in corridors.} Let us first briefly discuss the behavior of such agents in one-dimensional discrete corridors. In this case, the kinetic rules can be recast (in a continuum approximation) as a wave equation with density-dependent velocity (see End Matter), for which density solitons are solutions. Still, a key feature in open-ended and periodic corridors is that density fluctuations always tend to homogeneize. For instance, an initial density peak composed of agents similarly oriented in an otherwise empty and periodic corridor travels at a slowly decreasing speed while gradually vanishing (movie 2).

However, when a group of agents arrives at a dead end, the second kinetic rule tends to retain them until the incoming flow becomes weak, thus restoring a high density before the wave goes back. The direct consequence is that, in closed corridors, a density soliton can regenerate itself while bouncing at the boundaries. Actually, this effect is even strong enough to make solitons spontaneously emerge and self-maintain from an initial random distribution of the agents (fig.\ref{fig_1}-c, movie 3). These solitons have a comet-like density profile with a dense, fast-moving head and a long, slow-moving tail that develops in the intervals between rebounds at the boundaries.\\

\begin{figure}[b!]
  \includegraphics{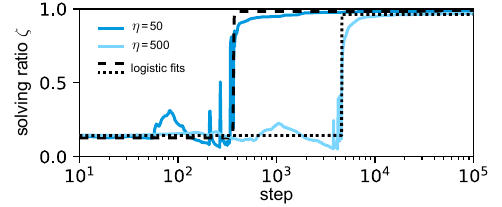}
  \caption{Traces of the solving ratio $\zeta$ (solid) for a maze of size $a=20$ (Prims) at $d=5$ with $\eta=50$ and $500$. Dashed curves are the corresponding logistic fits, from which the solving time $\tau$ is extracted.}
  \label{fig_2}
\end{figure}

\begin{figure*}[t!]
  \begin{minipage}[c]{0.67\textwidth}
    \includegraphics[width=0.95\textwidth]{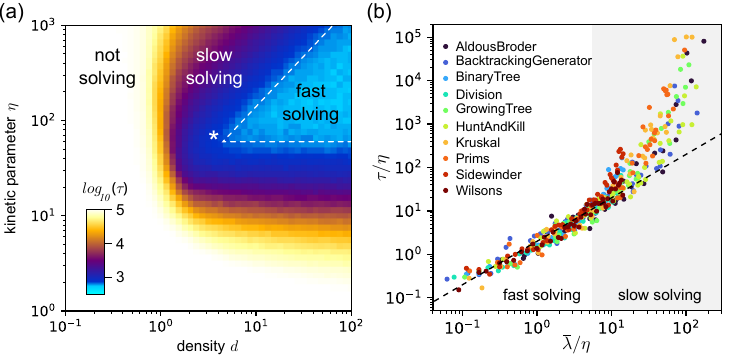}
  \end{minipage}\hfill
  \begin{minipage}[c]{0.33\textwidth}
    \caption{(a) Solving time $\tau$ as a function of $d$ and $\eta$ for mazes of size $a=20$ (Prims). At every grid position the times are averaged over $10^3$ runs ($100$ runs over $10$ mazes). The $\ast$ denotes the lowest density fast solving point.
    (b) Scaling of the solving time $\tau$ for 10 maze-generating algorithms. Each point stands for a single maze with a random size, a random density of agents and a random kinetic parameter $\eta$. The solving times are averaged over 10 runs. As reading guides, the dashed line is the $\tau=2\bar{\lambda}$ function and the grey zone indicates approximatively the transition from fast to slow solving.}
    \label{fig_3}
  \end{minipage}
\end{figure*}

% \subsection*{Behavior in mazes}
\label{sec:mazes}\textit{Behavior in mazes.} Let us now turn to a more complex geometry, \ie two-dimensional square mazes of side length $a$, such that $s=a^2$. Perfect mazes (\ie without loop) were generated with mazelib (see Methods) using 10 classical algorithms including Kruskal and Prims, whose references can be found in the mazelib package documentation. Additionally, we opened two walls at random on opposite sides to create entries. The \textit{solution} $S$ of the maze is then the unique path between the two entries, comprising $\lambda \in [a,s]$ squares. In order to keep the number of agents constant, we implemented periodic boundary conditions at these two entries; if the maze without entries is seen as a tree graph, this amounts to introducing an extra edge and forming a single loop of length $\lambda$. We chose Prims as a primary algorithm to showcase results since its solutions have the smallest $\lambda$ and the largest proportion of intersections, making the mazes \textit{a priori} more difficult to solve.

Due to the many dead ends in the mazes, swarms form with comet-like shapes reminiscent of the solitons in closed corridors (fig.\ref{fig_1}-d, movies 1 and 4). There are initially many swarms spread all across the maze, but they merge as they meet and occasionally split, ultimately leading to a low number of dense and long swarms.

Each swarm spontaneously explores local branches in a quasi-systematic manner: the second kinetic rule prevents it from immediately backtracking when it reaches an intersection since the outward flow in this direction is highly negative, giving a short-term persistence that forces exploration of a new branch. Then, the first kinetic rule allows agents in the tail to dramatically slow down, with a minimal average velocity of $v_{min}=1/(\eta+1)$ for isolated individuals. As agents keep their orientation while stalling, it creates an additional long-term persistence in the swarm tail by maintaining a minimal flow at every intersection. This is sufficient to guide incoming swarm heads in the same direction.

Now, if a swarm head comes upon its own tail, it means that it is evolving along a loop, hence on the solution path. Interestingly, the swarm then changes behavior and tends to homogenize its local densities, as observed in periodic and open-ended corridors. This confers strong stability and the swarm then stays in the solution virtually forever, a process termed here \textit{fixing}; smaller swarms entering in contact with this stable formation merge with it.

Such swarms consistently find the maze solution, in a scalable manner as we could observe robust solving from $a=3$ up to $a=1,000$ (movie 5), regardless of the maze generating algorithm.\\

% \subsection*{Solving ratio and parameter space}
\label{sec:param_space}\textit{Solving ratio and parameter space.}

Let us now quantify the success of swarm intelligence in our system by using the solving ratio, \ie the fraction of all agents that have reached the solution path:
\begin{equation}
  \zeta(t) = \frac{1}{N} \sum\limits_{i \in S} n_i(t)
  \label{eq:zeta}
\end{equation}

\noindent as a proxy for maze solving. It has values between $0$ and $1$, corresponding to no agent and all agents on the solution, respectively. Typical $\zeta$ traces for a sample maze are represented in fig.~\ref{fig_2} for two values of the kinetic parameter $\eta$. As observed here, pre-solving signatures are often similar for a given maze -- though stretched with $\eta$ -- but would differ for other mazes. We then fit $\zeta(t)$ with a logistic function
\begin{equation}
  f(t) = \zeta_0 + \frac{L}{1 + e^{-k(t-\tau)}}
  \label{eq:fit_zeta}
\end{equation}

\noindent to capture the moment where the swarm stabilizes on the solution and extract the corresponding solving time $\tau$ for each run. 
% Whenever the swarm fails to solve the maze, the fit fails and $\tau$ is not set. 

Then, we measured the average solving times in the $(d,\eta)$ parameter space (fig.~\ref{fig_3}-a). First, there is a transition from non-solving to solving sets of parameters. It is difficult to precisely estimate the curve where solving times diverge based on simulations with a limited number of iterations, and longer simulations may give a slightly shifted boundary. Yet, it is clear that such a transition exists, since the mazes cannot be solved and fixed when $d<\lambda/s$, \ie when there are fewer agents than the solution length.

Interestingly, fig.~\ref{fig_3}-a also reveals that there are two solving regimes, termed here \textit{slow} and \textit{fast} solving. Fast solving appears in a conic section far from the non-solving transition and is characterized by a constant and minimal average solving time all over the zone. In this regime, the swarm heads are sufficiently dense (\ie moving fast) and the tails sufficiently long to ensure they statistically meet during the initial exploration of all squares (see eq.~\ref{eq:scaling}).

Slow solving appears close to the non-solving transition border, and is due to the fact that slower swarm heads (low $d$, high $\eta$) or short tails (low $\eta$) decrease the probability that a swarm meets its own tail. This imposes a growing succession of correct choices by the swarm head at the intersections of the solution where there are no persistent agents, typically if they are already gone.\\

% \subsection*{Fast solving time scaling}
\label{sec:scaling}\textit{Fast solving time scaling.} We then tried to find how the solving time depends on the other two parameters, namely the generating algorithm and the maze size $s$. We generated hundreds of mazes of random sizes with 10 different algorithms, and ran simulations with random values of $d$ and $\eta$. We ensured that the density is always sufficient so that the upper-left slow solving regime of fig.~\ref{fig_3}-a is never reached. We then searched for the best scaling of $\tau$ as a function of the parameters and several elementary maze properties. We found that the number of squares outside the solution $\bar{\lambda} = s-\lambda$ is the best descriptor of the solving time, and $\tau/\eta$ is represented as a function of $\bar{\lambda}/\eta$ in fig.~\ref{fig_3}-b.

Several interesting observations can be made from this plot: first, neither the generating algorithm nor the density of agents seem to have an impact on the resolution time in this area of the parameter space. Then, the transition between the slow and fast regimes occurs at a fixed value of $\bar{\lambda}/\eta$, which is consistent with the straight horizontal demarcation line in fig.~\ref{fig_3}-a, though a given maze size corresponds to a distribution of $\bar{\lambda}$. Given a maze with a known solution length $\lambda$, a useful practical criterion is that the resolution should happen in the shortest time provided that:
\begin{equation}
  \eta \gtrsim \bar{\lambda}/5
  \label{eq:eta_criterion}
\end{equation}

Finally, it is remarkable that in the fast solving regime $\tau$ scales linearly with $\bar{\lambda}$ over two decades. The relation:
\begin{equation}
  \tau \simeq 2\bar{\lambda}
  \label{eq:scaling}
\end{equation}
\noindent provides a good approximation of the expected solving time. It is consistent with a very basic idea that all dead-end branches have to be explored once before the swarm locks the flow away from them at all the intersections of the solution path. Squares have then to be explored twice, for back and forth motion of the swarm head. This is of course an oversimplified view, since in practice there are several swarms exploring different parts of the maze at the same time, the swarm heads have a velocity always below 1 and some branches are explored several times, but it is striking to see that these effects globally even out in the fast solving regime.\\

% \section*{Discussion}
\label{sec:discussion}\textbf{Discussion.} In this Letter we introduce a model of agents obeying two simple but non-linear kinetic rules based on local densities distributed by orientations, and showed that it is sufficient to solve the task of solution finding and fixing in two-dimensional mazes. The emerging swarms show multiscale persistence abilities, with fast heads performing quasi-systematic local exploration and long tails keeping track of the current direction (exploitation). 

The results presented here can also be interpreted from a multi-objective optimization perspective. For systems where each agent has a cost, the quantities to minimize are the solving time, the density $d$ and the kinetic parameter $\eta$. For the latter indeed, when the solution is fixed there is still a small proportion of agents remaining isolated or in small groups that diffuse randomly at speeds as low as $v_{min}$; a smaller $\eta$ helps those stragglers reach the solution faster. A remarkable feature is that there exists a single Pareto-optimal point ($d^\ast$, $\eta^\ast=\bar{\lambda}/5$) in the parameter space. Moreover, in the fast solving regime, the time complexity is $O(s)$, indicating that these swarms are as efficient at solving mazes as many standard algorithms; for instance, the classical breadth-first search has the same time complexity but requires an additional explicit memorization of explored paths with space complexity $O(s)$~\cite{Cormen_2022}.

Going back to our initial question, this system defines a new kind of collective strategy for navigation tasks. It is not, however, the strategy employed by the vast majority of natural swarms. Reasons for this may be that the density of required active workers is relatively high, and that a few percent of the individuals may get lost in the maze for a very long time even after the solution is found. Natural colonies that can't afford this budget may have evolved more sophisticated strategies with memory, communication or stigmergy for instance, but these limitations may not be an issue for swarms of active particles or micro-robots, where individuals are usually cheap by design.

\bibliographystyle{apsrev4-1}
\bibliography{biblio}

% --------------------------------------------------------------------
%       END MATTER
% --------------------------------------------------------------------

\clearpage
\setcounter{equation}{0}
\setcounter{figure}{0}
\setcounter{table}{0}
\setcounter{page}{1}
\makeatletter
\renewcommand{\theequation}{S\arabic{equation}}
\renewcommand{\thefigure}{S\arabic{figure}}
\renewcommand{\thetable}{S\arabic{table}}
\setcounter{section}{0}
\renewcommand{\thesection}{S\arabic{section}}

\section*{End matter}

\subsection{Movies legends}

\textbf{Movie 1} Solving of a square maze of size $a=10$ by 200 agents with $\eta=25$, with random initial positions and orientations. Color indicates the number of agents on each square.

\textbf{Movie 2} Evolution of a density peak in a periodic corridor of size $a=30$, and estimation of the peak total displacement, indicating a constant speed.

\textbf{Movie 3} Evolution of the density in a closed corridor of size $a=30$, starting from a random distribution of agents. Density peaks spontaneously form and move at constant speed.

\textbf{Movie 4} Solving of a square maze of size $a=50$ by 50,000 agents ($d=20$) with $\eta=500$. Color indicates the number of agents on each square.

\textbf{Movie 5} Solving of a very large maze of size $a=1,000$ by $10^7$ agents ($d=10$) with $\eta=100,000$.

\subsection{Methods}

All the code used for this article has been written in Python and is available at the following repository: \url{https://github.com/CandelierLab/publi_maze}. We used mazelib (\url{https://pypi.org/project/mazelib/}) for generating the mazes, pyopencl (\url{https://pypi.org/project/pyopencl/}) to run on GPU and lib-anim (\url{https://pypi.org/project/lib-anim/}) for visualizations. All the maze generating algorithms used here are parameterless or use the default parameters. The only exception is \verb+GrowingTree+, for which we have set the parameter \verb+backtrack_chance+ to 0.5 so that it stands rigth in-between \verb+BackTrackingGenerator+ and \verb+Prims+.

\subsection{Corridor solitons}

Let us consider a group of agents in an open one-dimensional corridor and moving in the same direction. $n_i(t)$ is the density of agents at square $i$ at time $t$ and $p_i(t)$ their probability to move. For simplicity, $n_i(t)$ is assumed to be a continuous variable here. On square $i$, the density at $t+1$ is the sum of the agents that didn't move and the agents that moved from square $i-1$:
\begin{align}
& n_i(t+1) = n_i(t) \left( 1 - p_i(t) \right) + n_{i-1}(t) p_{i-1}(t) \\
& n_i(t+1) - n_i(t) = n_{i-1}(t) p_{i-1}(t) - n_i(t) p_i(t)
\end{align}

Writing simply $n$ the local density and $p$ the moving probability, let us define $f(n) = np(n) = \frac{n^2}{n+\eta}$. We have:
\begin{equation}
  \frac{\partial n}{\partial t} = - \frac{\partial f(n)}{\partial x}
\end{equation}

In the $\eta \gg n$ regime the approximation $f(n) \simeq n^2/\eta$ stands, and posing the change of variable $u=2n/\eta$ leads to the inviscid Burgers' equation~\cite{Burgers_1948}:
\begin{equation}
  \frac{\partial u}{\partial t} + u \frac{\partial u}{\partial x} =0
\end{equation}
\noindent which is a prototype for conservation equations developing discontinuities, also known as \textit{shock waves}.

More generally:
\begin{align}
  \frac{\partial^2 n}{\partial t^2} &= - \frac{\partial}{\partial x} \left( \frac{\partial f(n)}{\partial t} \right) = - \frac{d f}{d n} \frac{\partial}{\partial x} \left( \frac{\partial n}{\partial t} \right)\\
  \frac{\partial^2 n}{\partial t^2} &=  \left(\frac{df}{dn} \right)^2 \frac{\partial^2 n}{\partial x^2}
\end{align}
\noindent which is the one-dimensional d'Alembert wave equation, except that the wave celerity depends on the density $n$:
\begin{equation}
  c = \frac{df}{dn} = 1 - \left( \frac{\eta}{n+\eta} \right)^2
  \label{seq:c}
\end{equation}

This implies that clusters of agents with the same orientation move as solitons in the corridor. Eq.~\ref{seq:c} further indicates that larger peaks move faster and merge with smaller peaks, and that solitons moving to the right (resp. left) are negatively (resp. positively) skewed.

Then, the probabilistic nature of agent-based motion on a discrete mesh introduces some dispersion. Agents in the tail are statistically prevented from catching-up with the wave front and the tail progressively accumulates mass. As a result the wave amplitude slowly decreases and the wave eventually vanish, due to the discrete nature of density.

\end{document}